\documentclass[prd,groupedaddress,showpacs,showkeys,onecolumn,nofootinbib]{revtex4}
\usepackage{bm}
\begin{document}
\def\Barcelo{Barcel\'o}
\def\Eotvos{E\"otv\"os}
\title{Refringence, field theory, and normal modes}
\author{Carlos \Barcelo}
\email{carlos@hbar.wustl.edu}
\thanks{This work has been done at Washington University in Saint Louis
supported by the Spanish MCYT; Currently supported by
a European Community Marie Curie grant.}
\affiliation{Relativity and Cosmology group, School of Computer Science
and Mathematics, Portsmouth University, Portsmouth PO1 2EG, Britain.}
\author{Stefano Liberati}
\email{liberati@physics.umd.edu}
\homepage{http://www2.physics.umd.edu/~liberati}
\thanks{Supported by the US NSF}
\affiliation{Physics Department, University of Maryland, 
College Park, MD 20742--4111, USA}
\author{Matt Visser}
\email{visser@kiwi.wustl.edu}
\homepage{http://www.physics.wustl.edu/~visser}
\thanks{Supported by the US DOE}
\affiliation{Physics Department, Washington University, Saint
Louis MO 63130--4899, USA}
\vspace{1truecm}
\date{%
19 November 2001; 
\LaTeX-ed \today}
\bigskip
\begin{abstract}
In a previous paper [gr-qc/0104001; Class. Quant. Grav. {\bf 18}
(2001) 3595-3610] we have shown that the occurrence of curved
spacetime ``effective Lorentzian geometries'' is a generic result of
linearizing an arbitrary classical field theory around some
non-trivial background configuration. This observation explains the
ubiquitous nature of the ``analog models'' for general relativity that
have recently been developed based on condensed matter physics. In the
simple (single scalar field) situation analyzed in our previous paper,
there is a single unique effective metric; more complicated situations
can lead to bi-metric and multi-metric theories. In the present
paper we will investigate the conditions required to keep the
situation under control and compatible with experiment --- either by
enforcing a unique effective metric (as would be required to be
strictly compatible with the Einstein Equivalence Principle), or at
the worst by arranging things so that there are multiple metrics that
are all ``close'' to each other (in order to be compatible with the
{\Eotvos} experiment). The algebraically most general situation leads
to a physical model whose mathematical description requires an
extension of the usual notion of Finsler geometry to a
Lorentzian-signature pseudo-Finsler geometry; while this is possibly
of some interest in its own right, this particular case does not seem
to be immediately relevant for either particle physics or
gravitation. The key result is that wide classes of theories lend
themselves to an effective metric description. This observation
provides further evidence that the notion of ``analog gravity'' is
rather generic.
\end{abstract}
\maketitle

\def\half{{1\over2}}
\def\L{{\mathcal L}}
\def\S{{\mathcal S}}
\def\d{{\mathrm{d}}}
\def\etal{{\emph{et al.}}}
\def\det{{\mathrm{det}}}
\def\tr{{\mathrm{tr}}}
\def\ie{{\emph{i.e.}}}
\def\im{{\rm i}}
\def\bnabla{\mbox{\boldmath$\nabla$}}
\def\x{{\mathbf x}}
\def\aka{{\emph{aka}}}
\def\Choose#1#2{{#1 \choose #2}}
\def\etc{{\emph{etc.}}}
\def\HRULE{{\bigskip\hrule\bigskip}}
\section{Introduction}

Whenever you have a single classical scalar field governed by
Lagrangian dynamics, the linearization of the field around any
classical solution automatically provides you with a curved space
``effective metric''. Furthermore this effective metric has Lorentzian
signature if and only if the PDE governing the scalar field is
hyperbolic~\cite{normal-modes}. This observation, combined with the
fact that quantization of the linearized fluctuations automatically
provides an Einstein--Hilbert term in the one-loop effective action,
led us to suggest that Einstein gravity is a generic low-energy limit
of a wide class of quantum field theories~\cite{normal-modes}.

A technical step in~\cite{normal-modes} that kept the calculation
under control was to assume that one was dealing with a single scalar
field, which automatically implied that there was a single unique
effective metric. However, this assumption is very restrictive.  By
starting with a single scalar field it is not possible to reproduce
the entire algebraic structure of the set of possible metrics in
GR. Therefore, in the present article we turn our attention towards
the somewhat messier general situation of multiple fields (multiple
scalar fields, or a multi-component vector or tensor field). The key
new results in this situation are:
\begin{enumerate}
\item
In some situations the existence of a single unique effective metric
can be deduced. This is the case that is compatible with strict
application of the Einstein Equivalence Principle.
\item
More generally there may be several distinct effective metrics on the
same spacetime (this will generically occur if the dynamical system
exhibits ``birefringence'' --- more properly called
``multi-refringence'' --- a situation in which different types of
oscillation, typically referred to as different ``polarizations'', can
propagate at different speed). As long as these various metrics are
sufficiently ``close'' to each other, different types of matter will
feel approximately the same geometry --- as is required experimentally
to satisfy the constraints deduced from the {\Eotvos} experiment.
\item
In the algebraically most general situation a mathematical structure
is encountered which can best be viewed as a hyperbolic extension of
the notion of a Finsler geometry; these pseudo-Finsler geometries bear
the same relation to ordinary Finsler geometries that
pseudo-Riemannian geometries ({\aka} Lorentzian geometries) bear to
Riemannian geometries---and many of the same sort of technical
problems arise due to indefiniteness of the pseudo-Finslerian
``metric''. While these structures may be of interest in their own
right, they do not seem to be immediately appropriate for either
particle physics or gravitation.
\end{enumerate}
In our previous paper we studied the single-scalar case in detail;
that case being a perfect exemplar for the occurrence of a Lorentzian
effective metric~\cite{normal-modes}. That paper also contained an
extensive bibliography regarding analog models and we will be more
selective in this present paper.  We shall consider multi-component
systems and push the ``effective metric'' analysis as far as
practical, making extensive use of the theory of characteristic
surfaces. We derive algebraic conditions that should be satisfied to
keep the effective metric unique (or nearly so).  We investigate the
notion of ``polarization'' and the associated ``Fresnel equation'',
leading to the notion of ``multi-refringence''. Finally we have a few
words to say about the algebraically general pseudo-Finsler geometries
--- this is a subject that mathematically is very poorly developed,
and we hope by this paper to generate some interest in what otherwise
seems a rather arcane and abstract subject.  The key message to take
from the present effort is that while dealing with multiple fields is
algebraically more tricky, there are nevertheless wide classes of
dynamical systems that lend themselves to an effective metric
description. Furthermore with multiple background fields, the
background geometry is more flexible --- the metric could in principle
be completely general.

\section{Linearized fields from arbitrary background systems}
\label{SS:linearization}
\subsection{Lagrangian analysis}
\label{S:Lagrangian}

Suppose we consider a {\emph{collection}} of fields $\phi^A=\{\phi^1,
\phi^2,\dots\}$ whose dynamics is governed by some first-order
Lagrangian $\L( \partial_\mu \phi^A, \phi^A)$. Here ``first-order'' is
taken to mean that the Lagrangian depends only on the fields and their
first derivatives.  We want to consider linearized fluctuations around
some background solution of the equations of motion. As in the
single-field case~\cite{normal-modes} we write
\begin{equation}
\phi^A(t,\vec x) = \phi_0^A(t,\vec x) + \epsilon  \phi_1^A(t,\vec x) + 
{\epsilon^2\over2}  \phi_2^A(t,\vec x) + O(\epsilon^3).
\end{equation}
Now use this to expand the Lagrangian
\begin{eqnarray}
{\L}(\partial_\mu \phi^A,\phi^A) 
&=& 
{\L}(\partial_\mu \phi^A_0,\phi^A_0)
+\epsilon \left[ 
{\partial \L\over\partial(\partial_\mu \phi^A)} \; \partial_\mu \phi^A_1
+
{\partial \L\over\partial\phi^A} \; \phi^A_1
\right]
+
{\epsilon^2\over2} \left[ 
{\partial \L\over\partial(\partial_\mu \phi^A)} \; \partial_\mu \phi^A_2
+
{\partial \L\over\partial\phi^A} \;  \phi^A_2
\right]
\nonumber
\\
&+&
{\epsilon^2\over2} \Bigg[ 
{\partial^2 \L\over\partial(\partial_\mu \phi^A) \; \partial(\partial_\nu \phi^B)} 
\; \partial_\mu \phi^A_1 \; \partial_\nu \phi^B_1
+
2 {\partial^2 \L\over \partial(\partial_\mu \phi^A)\; \partial \phi^B} 
\; \partial_\mu \phi^A_1 \; \phi^B_1
+
{\partial^2 \L\over \partial\phi^A\; \partial\phi^B} 
\; \phi^A_1 \; \phi^B_1
\Bigg]
+
O(\epsilon^3).
\end{eqnarray}
Consider the action
\begin{equation}
S[\phi^A] = \int \d^{d+1} x \; \L(\partial_\mu\phi^A,\phi^A).
\end{equation}
Doing so allows us to integrate by parts. As in the single-field
case~\cite{normal-modes} we can use the Euler--Lagrange equations to
discard the linear terms (since we are linearizing around a solution
of the equations of motion) and so get
\begin{eqnarray}
S[\phi^A] &=& S[\phi_0^A] 
\nonumber
\\
&+&
{\epsilon^2\over2}
\int \d^{d+1} x \Bigg[
\left\{
  {\partial^2 \L\over
   \partial(\partial_\mu \phi^A) \; \partial(\partial_\nu \phi^B)} 
\right\}
\; \partial_\mu \phi^A_1
\; \partial_\nu \phi^B_1
+
2 
\left\{
{\partial^2 \L\over \partial(\partial_\mu \phi^A)\; \partial \phi^B} 
\right\}
\; \partial_\mu \phi^A_1 
\; \phi^B_1
+
\left\{
{\partial^2 \L\over \partial\phi^A\; \partial \phi^B} 
\right\}
\; \phi^A_1 \; \phi^B_1
\Bigg]
\nonumber
\\
&+&
O(\epsilon^3).
\end{eqnarray}
Because the fields now carry indices ($AB$\/) we cannot cast the
action into quite as simple a form as was possible in the single-field
case.  The equation of motion for the linearized fluctuations are now
read off as
\begin{eqnarray}
&&  \partial_\mu \left(\left\{
  {\partial^2 \L\over\partial(\partial_\mu \phi^A) \; \partial(\partial_\nu \phi^B)} 
   \right\}
 \partial_\nu \phi^B_1 \right)
+   \partial_\mu \left(
{\partial^2 \L\over \partial(\partial_\mu \phi^A)\; \partial \phi^B} 
\; \phi^B_1 \right)
-  \partial_\mu \phi^B_1 \; 
{\partial^2 \L\over \partial(\partial_\mu \phi^B)\; \partial \phi^A} 
- \left(
{\partial^2 \L\over \partial\phi^A\; \partial \phi^B} 
\right)
\phi^B_1
= 0. 
\label{E:unmaked-up}
\end{eqnarray}
This is a linear second-order {\emph{system}} of partial differential
equations with position-dependent coefficients. This system of PDEs is
automatically self-adjoint (with respect to the trivial ``flat''
measure $\d^{d+1} x$).

To simplify the notation we introduce a number of definitions. First
\begin{equation}
f^{\mu\nu}{}_{AB} \equiv
{1 \over 2}\left(
{\partial^2 {\cal L}\over\partial(\partial_\mu \phi^A) \; 
\partial(\partial_\nu \phi^B)}+
{\partial^2 {\cal L}\over\partial(\partial_\nu \phi^A) \; 
\partial(\partial_\mu \phi^B)} 
\right).
\end{equation}
This quantity is independently symmetric under interchange of $\mu$,
$\nu$ and $A$, $B$.  In the case of Bose fields we will want to
interpret this as something related to some sort of ``metric'', but
the interpretation is not as straightforward as for the single-field
case.

Next, define
\begin{equation}
\Gamma^\mu{}_{AB} \equiv
+
{\partial^2 {\cal L}\over \partial(\partial_\mu \phi^A)\; \partial \phi^B}
-
{\partial^2 {\cal L}\over \partial(\partial_\mu \phi^B)\; \partial \phi^A} 
+
{1 \over 2}\partial_{\nu}\left(
{\partial^2 {\cal L}\over\partial(\partial_\nu \phi^A) \; 
\partial(\partial_\mu \phi^B)}-
{\partial^2 {\cal L}\over\partial(\partial_\mu \phi^A) \; 
\partial(\partial_\nu \phi^B)} 
\right).
\end{equation}
This quantity is anti-symmetric in $A$, $B$. For Bose fields we will
want to interpret this as some sort of ``connexion''. Equivalently we
could write
\begin{equation}
\Gamma^\mu{}_{AB} \equiv
+
{\partial^2 {\cal L}\over \partial(\partial_\mu \phi^A)\; \partial \phi^B}
-
{\partial^2 {\cal L}\over \partial(\partial_\mu \phi^B)\; \partial \phi^A} 
-
{1 \over 2}\partial_{\nu}\left(
{\partial^2 {\cal L}\over\partial(\partial_\mu \phi^A) \; 
\partial(\partial_\nu \phi^B)}-
{\partial^2 {\cal L}\over\partial(\partial_\mu \phi^B) \; 
\partial(\partial_\nu \phi^A)} 
\right).
\end{equation}
It is useful to note that
\begin{equation}
\partial_\mu \; \Gamma^\mu{}_{AB} 
=
\partial_\mu \; {\partial^2 {\cal L}\over \partial(\partial_\mu \phi^A)\; 
\partial \phi^B}
-
\partial_\mu \; {\partial^2 {\cal L}\over \partial(\partial_\mu \phi^B)\; 
\partial \phi^A}.
\end{equation}
In the case of Fermi fields one typically has $f^{\mu\nu}{}_{AB} = 0$,
while the $ \Gamma^\mu{}_{AB}$ are most usefully thought of as
generalizations of the Dirac matrices.  Finally, define
\begin{equation}
K_{AB} =
-{\partial^2 \L\over \partial\phi^A\; \partial \phi^B} 
+
{1\over2}
\partial_\mu \left(
{\partial^2 \L\over \partial(\partial_\mu \phi^A)\; \partial \phi^B} 
\right)
+
{1\over2}
\partial_\mu \left(
{\partial^2 \L\over \partial(\partial_\mu \phi^B)\; \partial \phi^A} 
\right).
\end{equation}
This quantity is by construction symmetric in ($AB$\/). We will want
to interpret this as some sort of ``potential'' or ``mass matrix''.
Then the crucial point for the following discussion is to realize that
equation (\ref{E:unmaked-up}) can be written in the form
\begin{equation}
\partial_\mu \left( f^{\mu\nu}{}_{AB}\; \partial_\nu \phi^B_1 \right)
+ {1\over2} 
\left[ \Gamma^\mu_{AB} \; \partial_\mu \phi^B_1 
+ \partial_\mu (\Gamma^\mu_{AB} \; \phi^B_1) \right]
+ K_{AB} \; \phi^B_1 
= 0.
\label{E:system}
\end{equation}
Now it is more transparent that this is a formally self-adjoint
second-order linear {\em system} of PDEs.

\subsection{Systems of second-order hyperbolic PDEs}

One can arrive at a similar set of equations by directly linearizing
an arbitrary system of second-order PDEs, that is, without relying on
the existence of a Lagrangian dynamics.  Start by considering an
arbitrary system of second-order PDEs written in the form
\begin{equation}
G_A(x, \phi^B, \partial_\mu \phi^B, \partial_\mu \partial_\nu \phi^B) = 0.
\end{equation}
The PDE does not have to be linear or even quasi-linear; there are as
many equations as there are fields. As in the single-field case
defining hyperbolicity for such a general equation is not easy --- not
even Courant and Hilbert~\cite{Courant} consider this case
explicitly. Indeed we shall adapt their discussion and shall define
hyperbolicity in terms of the linearized equation: Suppose we
linearize around some solution $\phi_0^A$, writing
\begin{equation}
\phi^A(t,\vec x) = 
\phi_0^A(t,\vec x) + 
\epsilon  \phi_1^A(t,\vec x) + O(\epsilon^2).
\end{equation}
Then
\begin{equation}
{\partial G_A\over\partial(\partial_\mu \partial_\nu \phi^B)} \;
 \partial_\mu \partial_\nu \phi_1^B 
+
{\partial G_A\over\partial(\partial_\mu \phi^B)} \;
 \partial_\mu \phi_1^B 
+
{\partial G_A\over\partial\phi^B} \;
 \phi_1^B = 0.
\end{equation}
That is: the fluctuation satisfies a second-order system of linear
PDEs with time-dependent and position-dependent coefficients (these
coefficients depend on the background field you are linearizing
around).  This equation can be written in a manner similar to
(\ref{E:system})
\begin{equation}
\partial_\mu \left( \tilde f^{\mu\nu}{}_{AB}\; 
\partial_\nu \phi^{B}_1 \right)
+\tilde \Gamma^\mu_{AB} 
\; \partial_\mu \phi^{B}_1
+ \tilde K_{AB} \; \phi^{B}_1 
= 0,
\label{E:systempde}
\end{equation}
by defining
\begin{eqnarray}
&&\tilde f^{\mu\nu}{}_{AB}=
{\partial G_A\over\partial(\partial_\mu \partial_\nu \phi^B)},
\\
&&\tilde \Gamma^\mu_{AB}=
{\partial G_A\over\partial(\partial_\mu \phi^B)}-
\partial_\nu\left[
{\partial F_A\over\partial(\partial_\mu \partial_\nu \phi^B)}
\right],
\\
&&
\tilde K_{AB}={\partial G_A\over\partial\phi^B}.
\end{eqnarray}
Though the form of the coefficients is somewhat different from that
appearing in the Lagrangian-based analysis (in general the equation is
not automatically self-adjoint), much of the discussion that follows
can be applied (with minor modifications) to both cases.

\subsection{Eikonal approximation}

The theory of \emph{systems} of second-order PDE is relatively
complicated and much less transparent than that for a {\em single}
second-order PDE. (See, for example, Courant and
Hilbert~\cite{Courant}, volume 2, pp 577--618, or the Encyclopedic
Dictionary of Mathematics~\cite{EDM}.) We would like to be able to
construct some notion of spacetime metric directly from the quantities
$f^{\mu\nu}{}_{AB}$ (or $\tilde f^{\mu\nu}{}_{AB}$), but we shall soon
see that doing so is a somewhat tricky proposition.

Consider an eikonal approximation for an arbitrary direction
in field space, that is, take
\begin{equation}
\phi^A(x) = \epsilon^A(x) \; \exp[-i\varphi(x)],
\end{equation}
with $\epsilon^A(x)$ a slowly varying amplitude, and $\varphi(x)$ a
rapidly varying phase. In this eikonal approximation (where we neglect
gradients in the amplitude, and gradients in the coefficients of the
PDEs, retaining only the gradients of the phase) the linearized system
of PDEs (\ref{E:system}) becomes
\begin{equation}
\left\{
f^{\mu\nu}{}_{AB}\; \partial_\mu \varphi(x) \; \partial_\nu \varphi(x) 
+
\Gamma^\mu{}_{AB}\;   \partial_\mu \varphi(x)
+
K_{AB} 
\right\} \; 
\epsilon^B_1 = 0.
\label{E:eikonal}
\end{equation}
This has a nontrivial solution if and only if $\epsilon^A(x)$ is a
null eigenvector of the matrix 
\begin{equation}
f^{\mu\nu}{}_{AB}\; \partial_\mu \varphi(x) \; \partial_\nu \varphi(x) 
+ \Gamma^\mu{}_{AB}\;\partial_\mu \varphi(x) 
+ K_{AB}. 
\end{equation}
Now, the condition for such a null eigenvector to exist is that
\begin{equation}
F(p,q) \equiv 
\det \left\{ f^{\mu\nu}{}_{AB}\; 
\partial_\mu \varphi(x) \; \partial_\nu \varphi(x)
+
\Gamma^\mu{}_{AB}\;   \partial_\mu \varphi(x)
+
K_{AB}
\right\} = 0,
\label{E:fresnel}
\end{equation}
with the determinant to be taken on the field space indices. This is
the natural generalization to the current situation of the Fresnel
equation of bi-refringent optics~\cite{Born-Wolf,Landau}.  Following
the analogy with the situation in electrodynamics (either nonlinear
electrodynamics, or more prosaically propagation in a bi-refringent
crystal), the null eigenvector $\epsilon^A(x)$ would correspond to a
specific ``polarization''. The Fresnel equation then describes how
different polarizations can propagate at different velocities (or in
the language to be used later in the paper, can see different metric
structures).  In particle physics language this determinant condition
$F(p,q)=0$ is the natural generalization of the ``mass shell''
constraint. Indeed it is useful to define the mass shell as a subset
of the cotangent space by
\begin{equation}
{\cal F}(q) \equiv
\left\{ 
p_\mu \; \bigg| \; F(p,q) = 0
\right\}.
\label{E:mass-shell}
\end{equation}
In more mathematical language we are looking at the null space of the
determinant of the ``symbol'' of the system of PDEs. By investigating
$F(p,q)$ one can recover part (not all) of the information encoded in
the matrices $f^{\mu\nu}{}_{AB}$, $\Gamma^\mu{}_{AB}$, and $K_{AB}$,
or equivalently in the ``generalized Fresnel equation''
(\ref{E:fresnel}).  (Note that for the determinant equation to be
useful it should be non-vacuous; in particular one should carefully
eliminate all gauge and spurious degrees of freedom before
constructing this ``generalized Fresnel equation'', since otherwise
the determinant will be identically zero.  Some examples of this
phenomenon will be given later in the paper, see
subsection~\ref{S:spurious}).  We now want to make this analogy with
optics more precise, by carefully considering the notion of
characteristics and characteristic surfaces.  We will see how to
extract from the the high-frequency high-momentum regime described by
the eikonal approximation all the information concerning the causal
structure of the theory.

\section{Causal structures}
\label{S:causal}

One of the key structures that a Lorentzian spacetime metric provides
is the notion of causal relationships. This suggests that it may be
profitable to try to work backwards from the causal structure to
determine a Lorentzian metric. Now the causal structure implicit in
the system of second-order PDEs given in equation (\ref{E:system}) is
described in terms of the characteristic surfaces, and it is for this
reason that we now focus on characteristics as a way of encoding
causal structure, and as a surrogate for some notion of Lorentzian
metric. Note that via the Hadamard theory of surfaces of discontinuity
the characteristics can be identified with the infinite-momentum limit
of the eikonal approximation~\cite{Hadamard}. That is, when extracting
the characteristic surfaces we neglect subdominant terms in the
generalized Fresnel equation and focus only on the leading term in the
symbol ($f^{\mu\nu}{}_{AB}$). In particle physics language going to
the infinite-momentum limit puts us on the light cone instead of the
mass shell; and it is the light cone that is more useful in
determining causal structure.  The ``normal cone'' at some specified
point $q$, consisting of the locus of normals to the characteristic
surfaces, is defined by
\begin{equation}
{\cal N}(q) \equiv
\left\{ 
p_\mu \; \bigg| \; 
\det\left(f^{\mu\nu}{}_{AB} \;\; p_\mu \; p_\mu\right) = 0
\right\}.
\label{E:normal}
\end{equation}
As was the case for the Fresnel equation (\ref{E:fresnel}), the
determinant is to be taken on the field indices $AB$.  Remember to
eliminate spurious and gauge degrees of freedom so that this
determinant is not identically zero. (See Courant and
Hilbert~\cite{Courant}, volume 2 ``Partial differential equations'',
page 580.) We emphasise that the algebraic equation defining the
normal cone is the leading term in the Fresnel equation encountered in
discussing the eikonal approximation. If there are $N$ fields in total
then this ``normal cone'' will generically consist of $N$ nested
sheets each with the topology (not necessarily the geometry) of a
cone. Often several of these cones will coincide, which is not
particularly troublesome, but unfortunately it is also common for some
of these cones to be degenerate, which is more problematic. See
Courant and Hilbert footnote 1 on page 592:
\begin{quote}
{\sl It may be remarked that the present state of the theory of
algebraic surfaces does not permit entirely satisfactory applications
to the questions of reality of geometric structures which confront us
here.}
\end{quote}
Note that if one is dealing with Fermi fields where
$f^{\mu\nu}{}_{AB}=0$, one has a choice: either iterate the
first-order system of PDEs to produce a second-order system and
then apply the previous logic, or go to the highest nontrivial term
in the symbol and rephrase the discussion below ({\emph{mutatis
mutandis}}) in terms of
\begin{equation}
{\cal N}(q) \equiv
\left\{ 
p_\mu \; \bigg| \; 
\det\left(\Gamma^{\mu}{}_{AB} \;\; p_\mu \right) = 0
\right\}.
\label{E:normal-fermi}
\end{equation}

Returning to the Bosonic case, it is convenient to define a function
$Q(q,p)$ on the co-tangent bundle
\begin{equation}
Q(q,p) \equiv \det\left(f^{\mu\nu}{}_{AB}(q) \; p_\mu \; p_\mu \right).
\end{equation}
The function $Q(q,p)$ defines a completely-symmetric spacetime tensor
(actually, a tensor density) with $2N$ indices
\begin{equation}
Q(q,p) = Q^{\mu_1\nu_1\mu_2\nu_2\cdots\mu_N\nu_N}(q) 
\;\; p_{\mu_1} \; p_{\nu_1} \; p_{\mu_2} \; p_{\nu_2} \cdots \; 
p_{\mu_N} \;  p_{\nu_N}. 
\end{equation}
(Remember that $f^{\mu\nu}{}_{AB}$ is symmetric in both $\mu\nu$ and
$AB$ independently.)  Explicitly, using the expansion of the
determinant in terms of completely antisymmetric field-space
Levi--Civita tensors
\begin{eqnarray}
Q^{\mu_1\nu_1\mu_2\nu_2\cdots\mu_N\nu_N}(q) &=& {1\over N!} \;
\epsilon^{A_1 A_2 A_3 \cdots A_N} \; \epsilon^{B_1 B_2 B_3 \cdots B_N} \;
\\
&&
\qquad 
f^{\mu_1\nu_1}{}_{A_1 B_1} \;  f^{\mu_2\nu_2}{}_{A_2 B_2} \; 
\cdots  f^{\mu_N\nu_N}{}_{A_N B_N}.
\nonumber
\end{eqnarray}
In terms of this $Q(q,p)$ function the normal cone is
\begin{equation}
{\cal N}(q) \equiv
\left\{ 
p_\mu \; \bigg| \; Q(q,p) = 0
\right\}.
\end{equation}
In contrast, the ``Monge cone'' ({\aka} ``ray cone'', {\aka}
``characteristic cone'', {\aka} ``null cone'') is the envelope of the
set of characteristic surfaces through the point $q$.  Thus the
``Monge cone'' is dual to the ``normal cone'', its explicit
construction is given by (Courant and Hilbert~\cite{Courant}, volume
2, pp 583):
\begin{equation}
{\cal M}(q) = 
\left\{ 
t^\mu =  {\partial Q(q,p)\over\partial p_\mu} 
\;
\bigg| \; p_\mu \in {\cal N}(q)
\right\}.
\label{E:null}
\end{equation}
Equivalently
\begin{equation}
{\cal M}(q) = 
\left\{ 
t^\mu =  Q^{\mu\mu_2\cdots\mu_{2N}}(q)\;  
p_{\mu_2} \; p_{\mu_3} \cdots \; p_{\mu_{2N}} 
\;
\bigg| \; p_\mu \in {\cal N}(q)
\right\}.
\label{E:null2}
\end{equation}
Unfortunately, even if the normal cone is pleasantly behaved, the
Monge cone (null cone) may be quite messy. Indeed Courant and Hilbert
remark:
\begin{quote}
{\sl Even if [the normal cone] is a relatively simple algebraic cone
of degree [2N], the ray cone [Monge cone/null cone] may have
singularities, or isolated rays, and need not consist of separate
smooth conical shells.}
\end{quote}
(Courant and Hilbert~\cite{Courant}, volume 2, p 584). Defining the
notion of ``hyperbolicity'' in this general context boils down to the
question of just what constraints have to be placed on the original
system of PDEs to make sure the Monge cone is well-behaved.

For completeness, we mention a general construction for finding
bi-characteristic rays (these are lines which are guaranteed to lie on
characteristic surfaces). Start by picking some point $q$ and a
co-tangent vector $p$ which is in the normal cone ${\cal N}(q)$. Now
with these initial conditions solve the ``Hamiltonian equation''
\begin{equation}
{d q^\mu\over ds} = - {\partial Q(q,p)\over \partial p_\mu}; 
\qquad
{d p_\mu\over ds} = + {\partial Q(q,p)\over \partial q^\mu}.
\end{equation}
Here $s$ is just a parameter, it is not physical time. The resulting
curve [$q^\mu(s)$, $p_\mu(s)$] is called a bi-characteristic ray. For
all $s$ we have $p_\mu(s) \in {\cal N}(q(s))$ and $dq^\mu/ds \in {\cal
M}(q(s))$. (The momentum lies on the normal cone, and the velocity
lies on the Monge cone [null cone].)

In a manner similar to the way in which the Monge cone is the dual of the
normal cone, we can define a dual to the mass shell, this dual now
being a subset of the tangent space
\begin{equation}
{\cal F}^*(q) \equiv
\left\{ 
{\partial F(p,q)\over \partial p_\mu} \; \bigg| \; p \in {\cal F} (q)
\right\}.
\label{E:mass-shell2}
\end{equation}
In simple situations where there is a single metric the Monge cone can
be constructed simply by ``raising the index'' in the definition of the
normal cone, and similarly for the dual mass shell in terms of the
mass shell.

The structure of the normal and Monge cones encode all the information 
related with the causal propagation of signals associated with
the system of PDEs. In next section, we will detail how to relate
this causal structure with the existence of some spacetime
metrics in different specific situations, from the experimentally
favored single-metric theory deduced from the Equivalence Principle
to the most complicated case of pseudo-Finsler geometries.

\section{Geometrical interpretation}
\label{S:geometrical}
\subsection{Field redefinitions}

We have seen that the causal structure of a system of coupled PDEs of
the form (\ref{E:system}) could in general be rather
complicated. However, there are particular cases in which it is
relatively easy to find a geometrical interpretation of what is
happening.  First, it is important to realize that we are always free
to perform a field redefinition
\begin{equation}
\phi^A \to \bar\phi^A = h^A(\phi^B),
\end{equation}
essentially a coordinate change in field space. This induces a linear
transformation on the linearized fields $\phi^A_1$,
\begin{equation}
\phi^A_1 \to \bar\phi^A_1  
= 
\left.{\partial h^A\over\partial  \phi^B}\right|_{\phi^C_0} \; \phi^B_1
= 
[L^{-1}(\phi^C_0)]^A{}_B\;\; \phi^B_1,
\end{equation}
and, therefore, a redefinition of the $f^{\mu\nu}{}_{AB}$,
$\Gamma^\mu{}_{CD}$, and $K_{AB}$.  It is convenient to adopt matrix
notation, suppressing the field indices (but keeping the spacetime
indices explicit). In matrix notation
\begin{equation}
\bm{\phi}_1 \to \bm{\bar \phi}_1 = \bm{L^{-1}} \; \bm{\phi}_1,
\end{equation}
while the self-adjoint system of PDEs (\ref{E:system}) can be written
in the form
\begin{equation}
\partial_\mu \left( \bm{f}^{\mu\nu}\; \partial_\nu \bm\phi_1 \right)
+ \bm\Gamma^\mu \; \partial_\mu \bm\phi_1 
+ \half \partial_\mu (\bm\Gamma^\mu)  \; \bm\phi_1
+ \bm K \; \bm\phi_1
= 0.
\end{equation}
A brief computation yields
\begin{eqnarray}
\bm{f}^{\mu\nu} &\to& \bar{\bm{f}}^{\mu\nu} = 
\bm{L}^T \; \bm{f}^{\mu\nu} \; \bm{L};
\\
\nonumber
\\
\bm\Gamma^\mu &\to& \bar{\bm\Gamma}^\mu{} =
\bm{L}^T \; \bm\Gamma^\mu{} \;  \bm{L} 
+
\bm{L}^T \; \bm{f}^{\mu\nu} \; \partial_\nu \bm{L}
-
\partial_\nu \bm{L}^T \; \bm{f}^{\mu\nu} \; \bm{L};
\\
\nonumber
\\
\bm{K} &\to& \bar{\bm{K}} 
= 
\bm{L}^T \; \bm{K} \;  \bm{L} 
-\half\left(
\partial_\nu \bm{L}^T \; \bm{\Gamma}^{\nu} \; \bm{L}
-
\bm{L}^T \; \bm{\Gamma}^{\nu} \; \partial_\nu \bm{L}
\right)
+\half\partial_\mu \left(
\bm{L}^T \; \bm{f}^{\mu\nu} \; \partial_\nu \bm{L}
+
\partial_\nu \bm{L}^T \; \bm{f}^{\mu\nu} \; \bm{L}
\right)
\nonumber\\
&&
\qquad\qquad\qquad
- 
\partial_\nu \bm{L}^T \; \bm{f}^{\mu\nu} \; \partial_\mu \bm{L}.
\end{eqnarray}
Note that because of the matrix multiplications the ordering is
important. Also note that $(\bm{f}^{\mu\nu})^T =
\bm{f}^{\mu\nu}$, while $(\bm{\Gamma}^{\mu})^T =-
\bm{\Gamma}^{\mu}$ and $(\bm{K})^T = \bm{K}$, where the
transpose is on the field indices (not the spacetime indices), and
that these symmetry properties are preserved under these
background-field-dependent coordinate transformations.  The point is
that we can use these field redefinitions to simplify the matrix
$f^{\mu\nu}{}_{AB}$ using position-dependent general linear
transformations --- this will be useful below.  (For our purposes, the
detailed form of the non-homogeneous derivative terms is often not
important, though the fact of their presence is.)

\subsection{Einstein Equivalence Principle}

The physically simplest situation one can encounter is when the
multiple fields being analyzed simply correspond to different but
equivalent polarizations of a single field.  By looking at the eikonal
approximation of section \ref{S:Lagrangian}, one can verify if there
exist multiple independent polarizations that are solutions of
equation (\ref{E:eikonal}) for the {\emph{same}} phase function
$\varphi$. Those multiple solutions should be grouped into a single
class, all having the same propagation features. The number of
independent solutions grouped into equivalent classes will give us
information about the degree of degeneracy of the matrix
$f^{\mu\nu}{}_{AB}$.  (Here we are thinking of this object as a matrix
in field space, whose elements are spacetime matrices).

The simplest possible situation regarding the geometrical structure
emerging from a field theoretical mode analysis is that in which all
linearized fields can be grouped into a single class. This implies
that there must be some choice of field variables so that all the new
$\bar\phi^A_1$ see the same metric, that is:
\begin{equation}
\bar f^{\mu\nu}{}_{AB} = \delta_{AB} \;  f^{\mu\nu} =  
\delta_{AB} \;  \sqrt{-g} \; g^{\mu\nu}.
\end{equation}
This corresponds to the field theoretical analog PDE systems
obeying strict adherence to the Einstein Equivalence Principle. If we
use any other choice of field variables then we must have
\begin{equation}
f^{\mu\nu}{}_{AB} = h_{AB} \;  f^{\mu\nu} =  
h_{AB} \;  \sqrt{-g} \; g^{\mu\nu}.
\end{equation}
This ``factorization'' condition on $f^{\mu\nu}{}_{AB}$ is a necessary
and sufficient condition for strict adherence to the Einstein
Equivalence Principle. Indeed for strict adherence to the Einstein
Equivalence Principle you would also want to demand that in the
$\bar\phi^A_1$ field variables $\bar \Gamma^\mu{}_{AB} =0$, and one
would want the ``mass matrix'' $\bar K_{AB}$ to be position
independent. (At the very worst $\bar K_{AB}$ might contain curvature
coupling terms that would go to zero in the flat space limit.)

In the usual formulation of general relativity, strict adherence to
the Einstein Equivalence Principle is enforced by a policy of
``minimal substitution'' --- the matter Lagrangian is taken to be a
flat Minkowski-space Lagrangian with the substitution
$\eta_{\mu\nu}\to g_{\mu\nu}$. Since, by fiat, there is only one
metric put into this Lagrangian, a unique spacetime metric will emerge
for all the matter fields. (At least as long as all ``external
fields'' are set to zero.)  While the Einstein Equivalence Principle
is certainly compatible with experiment, strictly enforcing it may be
overkill --- indeed the low-energy field theories arising from string
theory and its relatives (in particular, various scalar tensor
theories) do not strictly adhere to the Einstein Equivalence
Principle, so it is useful to keep a little flexibility on this point.

\subsection{Multiple metrics}

Suppose now that we are looking for a multi-metric theory, then there
must be some choice of field variables so that all the linearized
fields $\bar\phi^A_1$ ``decouple'' and see independent metrics. That
is, there must be a ``diagonal'' representation in field space such
that
\begin{equation}
\bar f^{\mu\nu}{}_{AB} = 
{\mathrm{diag}}\{f^{\mu\nu}_1, f^{\mu\nu}_2, f^{\mu\nu}_3, 
\cdots f^{\mu\nu}_N\}
=  
\mathrm{diag}\{ \sqrt{-g_1}\; g^{\mu\nu}_1, \; \sqrt{-g_2}\; g^{\mu\nu}_2,  \;
\sqrt{-g_3}\; g^{\mu\nu}_3, \cdots  \sqrt{-g_N}\; g^{\mu\nu}_N\}.
\end{equation}
If we use any other choice of field variables then the necessary and
sufficient condition for the $ f^{\mu\nu}{}_{AB}$ to be simultaneously
diagonalizable in field space is that $\forall\;
\mu,\nu,\alpha,\beta$:
\begin{equation}
f^{\mu\nu}{}_{AB}\; f^{\alpha\beta}{}_{BC} = 
f^{\alpha\beta}{}_{AB}\; f^{\mu\nu}{}_{BC};
\qquad
\hbox{that is}
\qquad
[\; \bm{f}^{\mu\nu},  \bm{f}^{\alpha\beta} \,] = 0.
\end{equation}
Subject to this constraint, after diagonalizing $f^{\mu\nu}{}_{AB}$ in
field space, each field couples to its own independent spacetime
metric. If the original system of PDEs [equation (\ref{E:system})] is
hyperbolic then all these metrics will have Lorentzian signature.
Note that this is the behaviour that typically occurs in nonlinear
electrodynamics and in birefringent crystals; and note that such
birefringence is {\emph{not}} in conflict with the physical intent of
the Einstein Equivalence Principle because in both of these situations
there is an external field (the electromagnetic field; the rest frame
of the crystal itself) that softly breaks the Lorentz invariance.  One
thing that we can definitely say based on the {\Eotvos} experiment
[the observed universality of free fall] is that those low-energy
quantum fields relevant to describing ordinary bulk matter all see
approximately the same effective metric.

Note that the case of multiple metrics as defined above is related to
(though not identical to) the notion of ``reducibility'' as introduced
by Courant and Hilbert~\cite{Courant}, p 596.  The function $Q(q,p)$
is said to be ``reducible'' when it {\emph{factorizes}} into
lower-order polynomials. If it factorizes completely into simple
quadratic products then:
\begin{equation}
Q(q,p) = \prod_{A=1}^N Q_A(q,p) =  
\prod_{A=1}^N \left( h^{\mu\nu}_{A}(q) \; p_\mu \; p_\mu \right).
\end{equation}
If this ``quadratic reducibility'' property is satisfied then the
normal cone consists of $N$ nested topological cones each of which is
geometrically a cone, and we can define a variant ``multiple-metric
theory'' by defining the $N$ spacetime metrics using the prescription
\begin{equation}
\sqrt{-g_A} \;\;  g^{\mu\nu}_{A} = h^{\mu\nu}_{A}.
\end{equation}
Even then, in order for the system (\ref{E:system}) to be deemed to be
hyperbolic, one must additionally insist that {\emph{each}} of these
matrices have Lorentzian signature and that the causal structures
derived from each of these Lorentzian metric be compatible with each
other: not only should each Lorentzian metric satisfy some sort of
``chronology condition'' or ``causality condition'' (no closed null or
timelike paths), but to prevent an ill-posed problem one must insist
that one cannot even form closed causal loops by using a chain of
causal line segments belonging to different Lorentzian metrics.

Now if we have a multiple metric theory defined in terms of
simultaneously diagonalizing the $f^{\mu\nu}{}_{AB}$ this
automatically satisfies quadratic reducibility for $Q(q,p)$. The
converse is not necessarily true, and the class of multiple-metric
theories defined in terms of quadratic reducibility is more general
than that defined in terms of simultaneous diagonalization of the
kinetic terms. (This last comment is not supposed to be obvious; we
shall give an explicit example of this phenomenon when we discuss the
generic two-field situation in section~\ref{s:two-fields}.)

In many cases all these metrics will coincide, and we can speak of
{\em the} spacetime metric; this happens for instance (modulo some
technical issues such as gauge fixing) for both the Maxwell equation
and the Dirac equation. In terms of $Q(q,p)$ this requires
\begin{equation}
Q(q,p) = [Q_0(q,p)]^N =  
\left[\,h^{\mu\nu}_{0}(q) \; p_\mu \; p_\mu \,\right]^N.
\end{equation}
%

\subsection{The general case: pseudo-Finsler geometries}

The most general case, when $f^{\mu\nu}{}_{AB}$ cannot be described in
terms of a single metric, or even multiple metrics, must be dealt with
by the formalism of pseudo-Finsler geometries.  What we call
pseudo-Finsler geometries bear the same relation to Finsler geometries
that pseudo-Riemannian geometries ({\aka} Lorentzian geometries) bear
to Riemannian geometries---and they exhibit many of the same sort of
technical problems that arise due to indefiniteness of the ``metric''.

Remember that for Riemannian geometries all the physically interesting
quantities (metric, Riemann tensor, \etc) can be encoded in terms of
coincidence limits of the (real) distance function $d(x,y)$ and its
derivatives. If we try to extend this sort of analysis to
pseudo-Riemannian geometries then using the distance function is
awkward because it is sometimes real (spacelike separation), sometimes
zero (null separation), and sometimes pure imaginary (timelike
separation). Synge (and many others) have argued that the best way of
taking care of this technical difficulty is to define the ``world
function''~\cite{Synge}
\begin{equation}
\Omega(x,y) \equiv \half \; d(x,y)^2
\end{equation}
which is always guaranteed to be real (though it can be positive,
zero, or negative).

In the present situation we have found it useful to use the notion of
characteristic surfaces to encode the causal structure of the system
of PDEs given in equation (\ref{E:system}) and to define the real
quantity $Q(q,p)$ which can be positive, zero, or negative. Now
$Q(q,p)$ is not homogeneous linear in $p$, rather it is homogeneous of
order 2N:
\begin{equation}
Q(q,\lambda p) = \lambda^{2N} Q(q,p).
\end{equation}
If we try to define the analog of the normal Finsler distance
function, we would choose
\begin{equation}
d_F(q,p) = [Q(q,p)]^{1/2N}.
\end{equation}
If $Q(q,p)$ were always positive (which would correspond to an
elliptic system of PDEs) this construction can be used to provide a
real and positive distance function suitable for defining a Finsler
geometry.  Unfortunately, we are interested in the hyperbolic case, so
$Q(q,p)$ by definition possesses zeros and changes sign. So $d_F(q,p)$
is now generically complex; one typically encounters various branch
cuts involving 2N'th roots of $-1$, ($\exp\{i\pi/(2N)\}$), which invalidates
the standard presentation of Finsler geometries. (We emphasize that
this is a technical issue, not a fundamental issue, but it does mean
one cannot simply copy results from the standard mathematics
literature.)  Note that even in the case of a unique spacetime metric
(completely reducible, so that $Q(q,p) = [Q_0(q,p)]^N$) one still has
$d_F(q,p) = {\sqrt{Q_0(q,p)}} = d_0(q,p) $. In this case the Finsler
function degenerates to the usual Lorentzian distance distance
function (which is positive real, zero, or pure imaginary). In view of
the above, we see that it is $Q(p,q)$ itself that should be thought of
as fundamental: it is the natural pseudo-Finslerian generalization of
Synge's world function.

There have been several attempts at defining Lorentzian-signature
pseudo-Finsler geometries (for example, Asanov~\cite{Asanov}).
Unfortunately those pseudo-Finsler geometries are typically set up in
such a way as to {\emph{avoid}} the possibility of multiple light
cones, which is exactly the situation we are trying to probe in this
article. It does not seem to us that the Asanov formulation of
pseudo-Finsler geometries has anything to say about the situation at
hand.  While it is clear that in the general case we will want to
invoke some form of pseudo-Finsler geometry, none of the extant
formalisms are really
suitable~\cite{Bejancu,Chern,Rund,Antonelli,Chern2}. We hope that these
brief comments might stimulate some interest in further developing
this field.

\subsection{Hidden geometries --- Eliminating spurious fields}
\label{S:spurious}

There are cases in which straightforward application of the previous
analysis of characteristics does not work because the generalized
Fresnel equation (\ref{E:fresnel}) is identically zero.  This happens
in situations in which one of the fields is spurious, either because
of a gauge invariance or possibly because of some non-obvious
algebraic (not differential) relation between the fields. However, it
should be noted that in some of these cases one can eliminate the
spurious field completely, and thereby show that the remaining
physical fields are coupled to a ``reduced'' effective metric.

For example, let us take the general set of equations
(\ref{E:system}). Consider the situation in which for a particular
equation, say equation number $1$, and a particular field, say
$\phi_1^1$, we have $f^{\mu\nu}{}_{11}=0$, and $\Gamma^\mu{}_{11}=0$,
but $K_{11}\neq 0$. Now choose the notation $a,b$ to denote those
indices $A,B$ distinct from $1$. Then, the relevant first equation
from the system (\ref{E:system}) reads:
\begin{equation}
K_{11}\;\phi^1_1+
\partial_\mu \left( f^{\mu\nu}_{1{b}}\; 
\partial_\nu \phi^{b}_1 \right)
+ \half \left( 
\Gamma^\mu_{1{b}} \; \partial_\mu \phi^{b}_1
+
\partial_\mu [ \Gamma^\mu_{1{b}} \; \phi^{b}_1]
\right)
+ K_{1{b}} \; \phi^{b}_1 
= 0.
\end{equation}
This means that we can {\emph{algebraically}} solve for the field
$\phi_1^1$ using
\begin{equation}
\phi^1_1 = 
- 
{
\partial_\mu \left( f^{\mu\nu}_{1{b}}\; \partial_\nu 
\phi^{b}_1 \right)
+ \half\left(
\Gamma^\mu_{1{b}} \; \partial_\mu \phi^{b}_1
+
\partial_\mu [ \Gamma^\mu_{1{b}} \; \phi^{b}_1]
\right)
+ K_{1{b}} \; \phi^{b}_1 
\over K_{11}
}.
\end{equation}
This can now be used to eliminate $\phi_1^1$ from the system
(\ref{E:system}) leading to a reduced [$(N-1)\times(N-1)$] system of
equations.  To do this we first write the remaining $a$ equations as
\begin{equation}
\label{E:effeceq}
\partial_\mu \left( f^{\mu\nu}_{a{b}}\; \partial_\nu \phi^{b}_1 \right)
+\partial_\mu \left( f^{\mu\nu}_{a{1}}\; \partial_\nu \phi^{1}_1 \right)
+ \half \left( 
\Gamma^\mu_{a{b}} \; \partial_\mu \phi^{b}_1
+
\partial_\mu [ \Gamma^\mu_{a{b}} \; \phi^{b}_1]
\right)
+ \half \left( 
\Gamma^\mu_{a{1}} \; \partial_\mu \phi^{1}_1
+
\partial_\mu [ \Gamma^\mu_{a{1}} \; \phi^{1}_1]
\right)
+ K_{a{b}} \; \phi^{b}_1 
+ K_{a{1}} \; \phi^{1}_1 
= 0,
\end{equation}
and then substitute $\phi^1_1$. 

In general, this reduced set of equations has up to fourth-order
derivatives of the linearized fields and so cannot be analyzed along
the lines developed in this paper.  However, there exist some
particular situations in which the reduced set of equations is still
second order, though no longer formally self-adjoint.  Specifically,
within this method of eliminating spurious degrees of freedom, there
are three possible particular cases:
\begin{itemize}
\item
Case a: $f^{\mu\nu}{}_{1{b}}=0$ and $\Gamma^\mu_{1{b}}=0$,
\item
Case b: $f^{\mu\nu}{}_{{a}1}=0$ and $\Gamma^\mu_{{a}1}=0$,
\item
Case c: $f^{\mu\nu}{}_{1{b}}=0$ and $f^{\mu\nu}{}_{{a}1}=0$.
\end{itemize}
In all three of these cases the reduced set of equations can be
written in second-order form
\begin{equation}
\partial_\mu \left( \tilde f^{\mu\nu}{}_{a b}\; 
\partial_\nu \phi^{b}_1 \right)
+\tilde \Gamma^\mu_{a b} 
\; \partial_\mu \phi^{ b}_1
+ \tilde K_{a b} \; \phi^{b}_1 
= 0.
\label{E:redsystem}
\end{equation}
Specifically, we have:
\begin{itemize}
\item
Case a:
\begin{eqnarray}
&& \tilde f^{\mu\nu}_{{a}{b}}=
f^{\mu\nu}{}_{{a}{b}}
-f^{\mu\nu}{}_{{a}1}\;{ K_{1{b}} \over  K_{11} } 
\\
&& \tilde \Gamma^\mu_{{a}{b}}=
\Gamma^\mu_{{a}{b}}
-f^{\mu\nu}{}_{{a}1}
\;\partial_{\nu}\left({ K_{1{b}} \over  K_{11} } \right) 
-\Gamma^\mu_{{a}1}\;{ K_{1{b}} \over  K_{11} }
\\
&& \tilde K_{{a}{b}}=K_{{a}{b}}
-\partial_{\mu}\left[ f^{\mu\nu}{}_{{a}1}
\;\partial_{\nu} \left({ K_{1{b}} \over  K_{11} } \right)\right]
-\Gamma^\mu_{{a}1} 
\;\partial_{\mu}\left({ K_{1{b}} \over  K_{11} } \right)
-{1 \over 2}(\partial_{\mu}\Gamma^\mu_{{a}1})
\; { K_{1{b}} \over  K_{11} }
+{1 \over 2}(\partial_{\mu}\Gamma^\mu_{{a}{b}})
-K_{{a}1}\; { K_{1{b}} \over  K_{11} } ,
\end{eqnarray}
\item
Case b:
\begin{eqnarray}
&& \tilde f^{\mu\nu}_{{a}{b}}=
f^{\mu\nu}{}_{{a}{b}}
-f^{\mu\nu}{}_{1{b}}\; { K_{{a}1} \over  K_{11} }, 
\\
&& \tilde \Gamma^\mu_{{a}{b}}=
\Gamma^\mu_{{a}{b}}+
\partial_{\nu}
\left({ K_{{a}1} \over  K_{11} }\right)
\; f^{\nu\mu}{}_{1{b}}
-{ K_{{a}1} \over  K_{11} } \; \Gamma^\mu_{1{b}},  
\\
&& \tilde K_{{a}{b}}=
K_{{a}{b}}-
K_{{a}1} \; { K_{1{b}} \over  K_{11} }
-{1 \over 2}
(\partial_{\mu}\Gamma^\mu_{1{b}}){ K_{{a}1} \over  K_{11} }
+{1 \over 2}(\partial_{\mu}\Gamma^\mu_{{a}{b}}),
\end{eqnarray}
\item
Case c:
\begin{eqnarray}
&& \tilde f^{\mu\nu}_{{a}{b}}
=
f^{\mu\nu}{}_{{a}{b}}
-{1 \over 2K_{11}}\left(
\Gamma^\mu_{{a}1}\Gamma^\nu_{1{b}}
+\Gamma^\nu_{{a}1}\Gamma^\mu_{1{b}}
\right),
\\
&& \tilde \Gamma^\mu_{{a}{b}}
=
\Gamma^\mu_{{a}{b}}
+{1 \over 2K_{11}}(\partial_{\nu}\Gamma^\nu_{{a}1})
\Gamma^\mu_{1{b}}
-{1 \over 2K_{11}}\Gamma^\mu_{{a}1}
(\partial_{\nu}\Gamma^\nu_{1{b}})
+\partial_\nu\left[
{1 \over 2K_{11}}\Gamma^\mu_{{a}1}\Gamma^\nu_{1{b}}
-{1 \over 2K_{11}}\Gamma^\nu_{{a}1}\Gamma^\mu_{1{b}}
\right]
\nonumber\\
&&\hspace{1cm}
-{K_{1{b}} \over K_{11}}\Gamma^\mu_{{a}1}
-{K_{{a}1} \over K_{11}}\Gamma^\mu_{1{b}}
\\
&& \tilde K_{{a}{b}}
=
K_{{a}{b}}
-{1 \over 2}\Gamma^\mu_{{a}b}
\partial_{\mu}\left({1 \over K_{11}}\partial_{\nu}
\Gamma^\nu_{1{b}}\right)
-\Gamma^\mu_{{a}1}
\partial_{\mu}\left({K_{1{b}} \over K_{11}}\right)
-{1 \over 4K_{11}}(\partial_{\mu}\Gamma^\mu_{{a}1})
(\partial_{\nu}\Gamma^\nu_{1{b}})
\nonumber\\
&&\hspace{1cm}
-{K_{1{b}} \over 2K_{11}}(\partial_{\mu}\Gamma^\mu_{{a}1})
-{K_{{a}1} \over 2K_{11}}(\partial_{\mu}\Gamma^\mu_{1{b}})
-{K_{{a}1} \over K_{11}}K_{1{b}}
{1 \over 2}(\partial_{\mu}\Gamma^\mu_{{a}{b}}).
\label{casec}
\end{eqnarray}
\end{itemize}
If one now forgets about the initial non-linear system, the two cases
that follow the patterns (a) or (b) are actually trivial, in the sense
that the field $\phi^1_1$ can be seen as an artificial degree of
freedom introduced in order to write the equation of motion of a
$(N-1)$ fields as an $N$-component coupled system. Case (c) is a bit
more subtle. In fact, the Lagrangian of a irrotational barotropic
fluid follows this pattern. We will work out this example explicitly
in section~\ref{S:examples}.

Having completed the reduction, the new system of equations
(\ref{E:redsystem}) can be analyzed as to its causal and geometric
structures along the same lines as before. The new matrices $\tilde
f^{\mu\nu}{}_{{a}{b}}$ will determine the characteristics of the
reduced system.  Although the reduced system could now fail to be
self-adjoint, the previous analysis can nevertheless still be applied
--- this is because (once spurious fields have been eliminated) the
characteristics depend only on the leading symbol of the system of
PDEs and are insensitive to lower order terms; equivalently the
Fresnel equation determined form the eikonal approximation is
insensitive to any possible failure of self-adjointness. Once the
behaviour of the fields $\phi_1^{b}$ has been determined, the
remaining spurious field $\phi_1^{1}$ can (if desired) be obtained
from $\phi_1^{b}$ by differentiation.

\subsection{Hidden approximate geometries}
\label{SS:H}

Another situation of considerable interest occurs when, starting from
a system of $N$ coupled PDEs, one can neglect some of the degrees of
freedom in some particular limited regime. In this case one can again
perform a reduction process, but it is now important to remember that
the reduced system is only an approximation to the exact behaviour of
the system in the particular regimen.

Let us take units such that time and length have the same
dimension (some convenient reference velocity in the system is set to
be 1) and the elements of $K_{AB}$ are non-dimensional. Then, the
elements of $\Gamma^\mu_{AB}$ must have the dimension of length, and
those of the matrices $f^{\mu\nu}{}_{AB}$ the dimension of length
squared.

We can now assign to each dimensional coefficient in the set of
second-order differential equations (\ref{E:system}) a length scale.
For instance, thinking of case (c), we might have a situation in which
the length scales associated with the terms $f^{\mu\nu}{}_{11}$,
$\Gamma^\mu_{11}$, $f^{\mu\nu}{}_{1{b}}$, and $f^{\mu\nu}{}_{{a}1}$
are all much smaller than those of the remaining terms in the system
of equations.  Let $L_s$ be the largest length scale associated with
the terms $f^{\mu\nu}{}_{11}$, $\Gamma^\mu_{11}$,
$f^{\mu\nu}{}_{1{b}}$, and $f^{\mu\nu}{}_{{a}1}$, and let $L_l$ be the
smallest length scale among all the other terms. Then, provided $L_s
\ll L_l$, we can (in dealing with the physics at length scales larger than
$L_s$) safely approximate
\begin{equation}
f^{\mu\nu}{}_{11},\; \Gamma^\mu_{11}, \; f^{\mu\nu}{}_{1{b}},
\; f^{\mu\nu}{}_{{a}1} \simeq 0
\end{equation}
In these circumstances, for momenta low in comparison with $1/L_s$,
(large wavelengths in comparison with $L_s$) we will find that the
fields $\phi_1^{b}$ will behave as if they were coupled to an
approximate effective metric determined by
\begin{equation}
\tilde f^{\mu\nu}{}_{{a}{b}}=
       f^{\mu\nu}{}_{{a}{b}}
-{1 \over  2K_{11}} 
\left(\Gamma^\mu_{{a}1}\; \Gamma^\nu_{1{b}}
+ \Gamma^\nu_{{a}1}\; \Gamma^\mu_{1{b}}\right)
\end{equation}
This is exactly what happens with the effective Lorentzian metrics
arising in acoustic phenomena in dilute gas Bose--Einstein condensates
(BECs)~\cite{Barcelo,Garay}. In this case $L_s \sim h/(m\; c_0)$,
(where $h$ is Planck's constant, $m$ is the mass of the atoms making
up the gas, and $c_0$ is the velocity of sound in the
condensate). This means that the geometric acoustics approximation in
BECs is valid for wavelengths larger than $h/(m\;c_0)$ (an ``acoustic
Compton wavelength'') at which one would probe the discrete nature of
the condensed gas.

The characteristic surfaces that one can construct based in these
low-momentum metrics are only an approximation, and will only make
sense when exploring the system with momenta lower that the scale
$1/L_s$. By increasing the momentum we will find the true
characteristic surfaces of the system. For example, in a BEC the true
characteristics are those of a diffusion equation allowing infinite
speed propagation. Nevertheless the effective metric found in low
energies is a well defined Lorentzian metric with a maximum
propagation velocity~\cite{cpt01}.

\subsection{Fermionic fields}
\label{SS:Fermi}

The last possibility we shall comment on in this section is the
extreme case in which all the matrices $f^{\mu\nu}{}_{AB}$ are
strictly zero. When this happens we cannot directly apply the
previously derived analysis in terms of characteristic surfaces.  What
we should do instead is to either develop a modified discussion based
on definition (\ref{E:normal-fermi}), or more prosaically, to first
produce a system of second-order PDEs by iteration.  Apply the first
order operator
\begin{equation}
\hat D_{AB} \, \phi^B_1 \equiv
{1\over2} 
\left( \Gamma^\mu_{AB} \; \partial_\mu \phi^B_1 
+ \partial_\mu (\Gamma^\mu_{AB} \; \phi^B_1) \right)
+ K_{AB} \; \phi^B_1, 
\end{equation}
twice. Then
\begin{equation}
\hat D_{AC}\; \hat D_{CB} \, \phi^B_1=0.
\end{equation}
With this procedure we arrive at an equation of the form
\begin{equation}
\partial_\mu \left( \tilde f^{\mu\nu}{}_{AB}\; \partial_\nu \phi^B_1 \right)
+\tilde \Gamma^\mu_{AB} \; \partial_\mu \phi^B_1 
+ \tilde K_{AB} \; \phi^B_1 
= 0,
\label{E:fermisystem}
\end{equation}
with coefficients
\begin{eqnarray}
&& \tilde f^{\mu\nu}{}_{AB}=
{1 \over 2}(\Gamma^\mu{}_{AC}\;\Gamma^\nu{}_{CB}
+\Gamma^\nu{}_{AC}\;\Gamma^\mu{}_{CB}),
\\
&& \tilde \Gamma^\mu{}_{AB}=
K_{AC}\;\Gamma^\mu{}_{CB}+K_{BC}\;\Gamma^\mu{}_{CA}
+{1 \over 2}
\partial_{\nu}(\Gamma^\mu{}_{AC}\;\Gamma^\nu{}_{CB}
+\Gamma^\nu{}_{AC}\;\Gamma^\mu{}_{CB})
+(\partial_{\nu}\Gamma^\nu{}_{AC})\;\Gamma^\mu{}_{CB},
\\
&& \tilde K_{AB}=
\Gamma^\nu{}_{AC}\,\partial_{\nu}[(\partial_{\mu}\Gamma^\mu{}_{CB} )+K_{CB}]
+[(\partial_{\nu}\Gamma^\nu{}_{AC} )+K_{AC}]\,
[(\partial_{\mu}\Gamma^\mu{}_{CB} )+K_{CB}].
\end{eqnarray}
The system of equations (\ref{E:fermisystem}) now matches the pattern
of our previous discussion.

This situation is characteristic of the existence of fermionic
fields. For example in the trivial case of the Dirac equation in flat
space, we have $\Gamma^\nu{}_{AC}=\gamma^\nu{}_{AC}$ with $AB$ now
denoting spinorial indices. It is easy to see that
\begin{eqnarray}
&& \tilde f^{\mu\nu}{}_{AB}=
{1 \over 2}(\gamma^\mu{}_{AC}\;\gamma^\nu{}_{CB}
+\gamma^\nu{}_{AC}\;\gamma^\mu{}_{CB})
=\delta_{AB}\; \eta^{\mu\nu},
\end{eqnarray}
so that the different spinorial fields all feel the same Minkowski 
metric $\eta^{\mu\nu}$.

\section{Two--field systems}
\label{s:two-fields}

If we are dealing with a simple two-field system, much of the algebra
simplifies. This makes for a useful illustrative example. We start by
writing $Q(q,p)$ in terms of a completely symmetric four index tensor,
(a quartic)
\begin{equation}
Q(q,p) 
= 
Q(q)^{\mu\nu\rho\sigma} \; p_\mu \; p_\mu \; p_\rho \; p_\sigma
=
\left\{
f^{\mu\nu}{}_{11}(q) \; f^{\rho\sigma}{}_{22}(q) 
- 
f^{\mu\nu}{}_{12}(q) \; f^{\rho\sigma}{}_{12}(q) 
\right\}     
\; p_\mu \; p_\mu \; p_\rho \; p_\sigma.
\label{E:quartic}
\end{equation}
In discussing the most general causal structure for the propagation of
a two-field system it is this quartic $Q(q)^{\mu\nu\rho\sigma}$
that is the main geometrical object; instead of the metric
$f(q)^{\mu\nu}=\sqrt{-g}\; g^{\mu\nu}$ as in the single-field
case~\cite{normal-modes}. We can think of $f^{\mu\nu}{}_{AB}$ as a
$2\times 2$ matrix in field space, $f^{\mu\nu}{}_{AB}$, whose
components are point-dependent tensors (tensor densities, in fact):
\begin{equation}
f^{\mu\nu}{}_{AB} = 
\left[\begin{array}{cc}
f^{\mu\nu}{}_{11}&f^{\mu\nu}{}_{12}\\
f^{\mu\nu}{}_{12}&f^{\mu\nu}{}_{22}
\end{array}\right].
\end{equation}
By redefining the linearized fields we can transform
\begin{equation}
f^{\mu\nu}{}_{AB} \to \bar f^{\mu\nu}{}_{AB} = 
L_A{}^C \; L_B{}^D \; f^{\mu\nu}{}_{CD} \equiv 
M_{AB}{}^{CD}  \; f^{\mu\nu}{}_{AB}.
\end{equation}
In general, there is no combination of the three independent tensor
components of $f^{\mu\nu}{}_{AB}$, even with point-dependent
coefficients, that vanishes. That is, typically
\begin{equation}
M_{AB}{}^{11}(x)\;f^{\mu\nu}{}_{11}(x)+
M_{AB}{}^{22}(x)\;f^{\mu\nu}{}_{22}(x)+
2 M_{AB}{}^{12}(x)\;f^{\mu\nu}{}_{12}(x)\neq 0.
\end{equation}
In this algebraically most general case, we are forced to deal with
pseudo-Finsler geometries.  There are other cases in which the three
tensor components of $f^{\mu\nu}{}_{AB}$ are not independent. Then, by
using field redefinitions we can arrive at five different canonical
cases:
\begin{itemize}

\item[I:] Two independent components:

\begin{itemize}

\item[Ia:] A diagonal component is set to zero ---
\begin{equation}
f_{AB}=
\left[ \matrix{ f_{11} & f_{12} \cr
                f_{12} & 0      \cr } \right].
\end{equation}
In this case
\begin{equation}
Q(q,p) = - \left(  f^{\mu\nu}{}_{12}\; p_\mu \; p_\nu \right)^2.
\end{equation}
This case corresponds to quadratic reducibility for the function
$Q(q,p)$ \emph{without diagonalization of the kinetic terms}. There
are two induced metrics, both based on $f^{\mu\nu}{}_{12}= \sqrt{-g}\;
g^{\mu\nu}$, which identical to each other.  However these induced
metrics, while they successfully reproduce $Q(q,p)$, and so reproduce
the characteristic surfaces, are not enough to reproduce the symbol of
the system of PDEs. From the induced metrics we can reconstruct
$Q(q,p)$ but not $f^{\mu\nu}{}_{AB}$, so that we cannot arrange to
diagonalize the system of PDEs in field space.

\item[Ib:] The off-diagonal component is zero ---
\begin{equation}
f_{AB}=
\left[ \matrix{ f_{11} &    0       \cr
                0      & f_{22}     \cr } \right].
\end{equation}
In this case
\begin{equation}
Q(q,p) = \left(  f^{\mu\nu}{}_{11}\; p_\mu \; p_\nu \right) \; 
         \left(  f^{\mu\nu}{}_{22}\; p_\mu \; p_\nu \right).
\end{equation}
It is this case that corresponds to a proper bi-metric theory with
diagonal kinetic terms and two distinct metrics.
\end{itemize}

\item[II:] There is only one independent component: 

\begin{itemize}

\item[IIa:]
Only one diagonal component is different from zero ---
\begin{equation}
f_{AB}=
\left[ \matrix{ f_{11} &    0       \cr
                0      &    0       \cr } \right].
\end{equation}
In this case $Q(q,p)\equiv 0$; this is a sign that one of the fields
is unphysical, either because it is superfluous or because it is
actually a gauge degree of freedom.
\item[IIb:] 
The two diagonal components are different from zero ---
and equal
\begin{equation}
f_{AB}=
\left[ \matrix{ f_{11} &    0       \cr
                0      &    f_{11}      \cr } \right].
\end{equation}
This is a true single-metric theory with diagonalizable kinetic energy
terms and
\begin{equation}
Q(q,p) = - \left(  f^{\mu\nu}{}_{11}\; p_\mu \; p_\nu \right)^2.
\end{equation}
This is the case compatible with strict adherence to the Einstein
Equivalence Principle.

\end{itemize}

\item[III:] All the tensor components of $f_{AB}$ vanish:
\\ 
This probably means you are dealing with a Fermi field and somehow did
not notice. See section (\ref{SS:Fermi}) above.
\end{itemize}
In the three cases $I_a$, $I_b$ and $I\!I_b$ the quartic
(\ref{E:quartic}) factorizes into the product of two quadratics. The
case $I_b$ corresponds to a proper bi-metric theory in which there are
two fields each reacting to a different metric. In the cases $I_a$ and
$I\!I_b$ the two fields feel the same metric and so we would recover
the usual behaviour of classical fields (such as Maxwell or Dirac
fields) in general relativity.  Of course in all these cases it is
still necessary to analyze if the relevant metrics are
Lorentzian. (And to verify that they satisfy suitable causality
constraints.)  The case $I\!I_a$ is more tricky: The quartic
(\ref{E:quartic}) is degenerate and so we cannot strictly define any
kind of characteristic surface. However, as we have previously argued,
in some circumstances it is still possible to reduce the number of
fields and so define characteristic surfaces based in the existence of
a hidden metric.  We shall now demonstrate this and related phenomenon
explicitly with a pair of simple examples.

\section{Examples}
\label{S:examples}

\subsection{The barotropic irrotational inviscid fluid}
We can write the Lagrangian for a barotropic inviscid irrotational
fluid as~\cite{Jackiw,Raifeartaigh}
\begin{equation}
{\cal L}={1 \over 2}\rho (\bnabla \theta)^2+\rho \;\partial_t \theta
+\int_0^\rho d\rho' \; h(\rho').
\label{E:barotropic}
\end{equation}
Here, the two fields $\phi^A\equiv(\rho,\,\theta)$ are the fluid density
$\rho$ and the velocity potential $\theta$ (the velocity field ${\bf v}$
can be written as ${\bf v}=\bnabla \theta$ because the irrotational nature
of the fluid considered).  The function
\begin{equation}
h(\rho)=h[p(\rho)]=\int_0^{p}\;{dp' \over \rho(p')}
\end{equation}
is the enthalpy of the fluid.  Now, by linearizing we obtain the
relevant second-order Lagrangian for the perturbations
\begin{equation}
{\cal L}^{(2)}={1 \over 2}\rho_0 (\bnabla \theta_1)^2+
\rho_1 \bnabla \theta_1 \cdot \bnabla \theta_0 
+\rho_1 \partial_t \theta_1 
+{1 \over 2\rho_0} {dp \over d\rho}\bigg|_0 \; \rho_1^2.
\end{equation}
(Remember that the linear terms cancel by using the background
equation of motion).  From this we get the two linearized equations of
motion
\begin{eqnarray}
&& 
-\left(\partial_t \theta_1 +{\bf v_0} \cdot \bnabla \theta_1
+{c^2_0 \over \rho_0} \,\rho_1\right)=0,   
\label{eq1} \\
&& 
\partial_t \rho_1 + \bnabla(\rho_0 \bnabla \theta_1)+
\bnabla( \rho_1 {\bf v_0} )=0, 
\label{eq2}
\end{eqnarray}
in which we have used the definitions ${\bf v_0} \equiv \bnabla
\theta_0 $ and $c^2_0 \equiv (dp /d\rho)|_0$. (Of course, these must
and do agree with the equations obtained by the more traditional means
of first finding the full equations of motion and then linearizing.)
As we can see, equation (\ref{eq1}) has only the zero-order term for
$\rho_1$ and in the discussion of hidden geometries the condition {\it
(c)} above is fulfilled.  In our previous notation (note that
$\bm\rho$ and $\bm\theta$, which we boldface for clarity, are now
field indices not spacetime indices)
\begin{eqnarray}
&&
f^{\mu\nu}{}_{\bm{\rho\rho}} = f^{\mu\nu}{}_{\bm{\rho\theta}} = 
f^{\mu\nu}{}_{\bm{\theta\rho}} = 0,
\\
&&
f^{\mu\nu}{}_{\bm{\theta\theta}}=
\left[ \matrix{0 & \vdots & 0 \cr
               \cdots & \cdot & \cdots \cr
               0 & \vdots & \rho_0\;  \delta^{ij} \cr } \right],
\end{eqnarray}
and
\begin{eqnarray}
&& \Gamma^\mu{}_{\bm{\rho\rho}}= \Gamma^\mu{}_{\bm{\theta\theta}}= 0, \\
&& \Gamma^\mu{}_{\bm{\theta\rho}}= \{1, v_0^i  \}, \\
&& \Gamma^\nu{}_{\bm{\rho\theta}}=\{-1, -v_0^i  \},
\end{eqnarray}
while
\begin{eqnarray}
&& K_{\bm{\theta\theta}} = K_{\bm{\rho\theta}}= K_{\bm{\theta\rho}} = 0,\\
&& K_{\bm{\rho\rho}}=-{c^2_0 \over \rho_0}.
\end{eqnarray}
Finally, by looking at case (c) as given in equation (\ref{casec}) we
can calculate all the terms in equation (\ref{E:redsystem}).  We have
$\tilde K=0$, $\tilde \Gamma=0$ and a {\emph{single}} reduced metric
of the form
\begin{equation}
{\tilde f}^{\mu\nu}={\rho_0 \over c^2_0}
\left[ \matrix{-1 & \vdots & -v_0^i \cr
               \cdots & \cdot & \cdots \cdots \cdots \cdots \cr
               -v_0^i & \vdots & c^2_0 \delta^{ij} -v_0^i\; v_0^j \cr } \right],
\end{equation}
which corresponds to the usual acoustic metric\cite{Unruh,Visser}.
Once we have the solutions for $\theta_1$, the solutions for $\rho_1$
can be found by substituting in (\ref{eq1}).  

On the other hand, if we tried to eliminate $\theta_1$ to find an
equation from which to derive the same set of solutions for $\rho_1$,
we would find a very involved equation. Formally one can write
\begin{equation}
\rho_1=\Gamma^{\mu}{}_{\bm{\rho\theta}} \; \partial_{\mu} \theta_1={\cal D}\;\theta_1,
\end{equation}
which implies
\begin{equation}
\theta_1 ={\cal D}^{-1} \rho_1.
\end{equation}
Then we have 
\begin{equation} 
\partial_{\mu} (f^{\mu\nu} \partial_{\nu} \theta_1)=0,
\end{equation}
and 
\begin{equation}
{\cal D}\partial_{\mu} (f^{\mu\nu} \partial_{\nu} {\cal D}^{-1} \rho_1)=0.
\end{equation}
While at low momentum the differential equation satisfied by $\rho_1$
is rather complicated, we can see that in the large momentum limit
(the eikonal approximation) the propagation of density waves sees the
same causal structure as the $\theta_1$-waves. That is, the causal
structure for the propagation of both fields is described by the same
metric determined by $f^{\mu\nu}=\sqrt{-g}\; g^{\mu\nu}$.

Of course with hindsight the reason this elimination procedure works
is that it is possible to go all the way back to the Lagrangian in
equation (\ref{E:barotropic}) and eliminate the density from the
Lagrangian before linearizing. Nevertheless, this is a nice simple
example of how to eliminate spurious fields from the linearized
system.

\subsection{The BEC analog model}

Let us consider now a Bose--Einstein condensate.  It can be described
by the Gross--Pitaevskii equation,
\begin{equation}
\label{E:tdlg}
 \im \hbar \; \frac{\partial }{\partial t} \psi(t,\x)= \left (
 - {\hbar^2\over2m}\nabla^2 
 + V_{\rm ext}(\x)
 + \lambda \; \left| \psi(t,\x) \right|^2 \right) \psi(t,\x),
\label{eq:GP}
\end{equation}
where $\psi(t,\x)$ represents the mean field wave function.
By using the Madelung representation~\cite{Madelung}
\begin{equation}
\psi(t,\x) = \sqrt{\rho(t,\x)} \; \exp[-\im m \theta(t,\x)/\hbar],
\end{equation}
one can separate the (complex) Gross-Pitaevskii equation into the two
(real) equations:
\begin{equation}
\partial_t \theta + {1\over 2}(\bnabla \theta)^2 
+ {V_{\rm ext} \over m}
+ {\lambda \over m}\rho
- {\hbar^2 \over 2m^2}\; 
\left( {\Delta \sqrt\rho\over\sqrt\rho} \right)= 0.
\label{E:HJ}
\end{equation}
\begin{equation}
\partial_t \rho + \bnabla \cdot (\rho \; \bnabla \theta) = 0.
\label{E:continuity}
\end{equation}
Here, the quantity  
\begin{equation}
 V_{\rm Q}(\rho) \equiv  -  {\hbar^2 \over 2m} 
 \left( {\Delta \sqrt\rho\over\sqrt\rho}\right)
\end{equation}
is what is commonly called the ``quantum
potential''~\cite{Bohm,Holland}.  Expanding around a background
solution of these equations one finds the following two equations for
the perturbed quantities $\rho_1$ and $\theta_1$
\begin{eqnarray}
-\partial_t \theta_1  
-  {\bf v_0} \cdot \bnabla \theta_1 
- {c^2_0 \over \rho_0} \rho_1 + {\hbar^2\over 2 m^2}\; D_2 \rho_1 = 0.
\label{eqwithqp}
\end{eqnarray}
\begin{eqnarray}
\partial_t \rho_1 + \bnabla(\rho_0 \bnabla \phi_1) +
\bnabla( \rho_1 {\bf v_0} )=0
\end{eqnarray}
We have used the definitions ${\bf v_0} \equiv \bnabla \phi_0$ and
$c_0^2 \equiv (\lambda/m)\rho_0$. Additionally, $D_2$ represents a
relatively messy second-order differential operator obtained from
linearizing the quantum potential. Explicitly:
\begin{eqnarray}
D_2 \,\rho_1 
\equiv
-\half  \rho_0^{-3/2} \;[\Delta (\rho_0^{+1/2})]\;  \rho_1
+\half  \rho_0^{-1/2} \;\Delta (\rho_0^{-1/2} \rho_1).
\end{eqnarray}
These linearized equations are formally equivalent to the ones
obtained for the barotropic irrotational fluid except for the presence
of the quantum potential.  This difference amounts to the existence of
the following additional non-zero coefficients:
\begin{eqnarray}
&&
f^{\mu\nu}{}_{\bm{\rho\rho}}={\hbar^2\over 2 m^2}
\left[ \matrix{0 & \vdots & 0 \cr
               \cdots & \cdot & \cdots \cr
               0 & \vdots & {1 \over 2}\rho_0^{-1} \delta^{ij} \cr } \right],
\label{newf}
\\
&&
\Gamma^{\mu}{}_{\bm{\rho\rho}}= {\hbar^2\over 2 m^2} \bigg\{0,\;
\rho_0^{-1/2} \partial^i \rho_0^{-1/2} \bigg\},
\label{newg}
\\
&&
K_{\bm{\rho\rho}}={\hbar^2\over 2 m^2}\left[- \rho_0^{-3/2} 
\;\Delta (\rho_0^{+1/2})
+\rho_0^{-3/2}\bnabla \rho_0^{+1/2}\cdot \bnabla 
\rho_0^{+1/2}\right].
\label{newk}
\end{eqnarray}
Now, following the discussion in the ``Hidden approximate geometries''
subsection \ref{SS:H}, one can make the $K$ coefficient in equation
(\ref{eqwithqp}) non-dimensional by multiplying the whole equation by
$\rho_0/c_0^2$. Then, the important point from our example is that it
is easy to see that the new terms that occur in the BEC system (with
respect to the previous fluid system) all come naturally multiplied by
the the length scale $L_s=\hbar /(m\;c_0)$. This corresponds to an
``acoustic Compton wavelength'' defined by the propagation speed of
phonons in the condensate.  If one is probing the BEC system at large
length scales (low momenta) in comparison with this Compton
wavelength, one can neglect the new terms (\ref{newf}), (\ref{newg}),
and (\ref{newk}).  One can then perform the same trick as with the
barotropic irrotational fluid to obtain an effective metric for
$\rho_1$.  If we don't neglect the new terms coming from the quantum
potential we can still arrive at an integral-differential equation for
$\rho_1$. This equation will encode the Bogoliubov dispersion
relation~\cite{Barcelo,Garay}, showing us that the symbol of the
Gross--Pitaevskii equation is that of a (complex) diffusion equation.
As such it allows the propagation of signals at arbitrarily large
speeds~\cite{cpt01}.

\section{Summary and Discussion}

Taking an arbitrary system of hyperbolic second-order PDEs (either
generic or arising from a first-order Lagrangian), the behaviour of
the perturbations of the fundamental fields around any background
configuration can be given a geometrical interpretation.  For a single
field there is always a nice and clean geometrical interpretation in
terms of the d'Alembertian operator in an effective (typically curved)
Lorentzian geometry~\cite{normal-modes}.  For several coupled fields,
as discussed in this article, the situation is seen to be more
complex: In simple cases all the fields see the same effective
Lorentzian metric; with a little less luck the fields see different
effective Lorentzian metrics, up to one effective metric per field;
with extremely bad luck you will need to adopt some Finsler-like
extension of the notion of Lorentzian geometry.  Taking into account
this formal analysis, in order to obtain an analog model of general
relativity from a general system of PDEs it will be necessary to first
require that the system fulfill some sort of ``Einstein Equivalence
Principle'' (at least approximately). That is, at low energies we
would desire an (approximate) factorization:
\begin{equation}
f^{\mu\nu}{}_{AB} \approx h_{AB} \;  f^{\mu\nu} =  
h_{AB} \;  \sqrt{-g} \; g^{\mu\nu}.
\end{equation}
Then, and only then, will all the low-energy fields see
(approximately) the same effective Lorentzian metric, as is
experimentally implied by the {\Eotvos} experiment. At this level, we
have not found it possible to obtain the ``Einstein Equivalence
Principle'' from more fundamental principles.

In searching for a geometrical interpretation coming from a set of
PDEs we have analyzed the physical content of the matrix $f_{AB}$.
Sometimes, however, this matrix is singular and does not provide a
geometrical interpretation for the behaviour of all of the fields. In
these situations one can still find such geometrical interpretation by
extracting additional information from the $\Gamma^{\mu}{}_{AB}$ and
$K_{AB}$ coefficients. An extreme example is provided by a Fermi
system.  In this case, all the $f^{\mu\nu}_{AB}$ are zero but we can
find a geometrical structure coming from the $\Gamma^{\mu}{}_{AB}$. In
other situations, it can happen that the system has some ``spurious''
degrees of freedom making difficult the search for a geometrical
interpretation in terms of the $f^{\mu\nu}{}_{AB}$ alone.  In many
situations we can eliminate these spurious degrees of freedom and
arrive to a reduced set of PDEs with a well defined geometrical
interpretation. Additionally, there can be situations in which the
elimination of spurious degrees of freedom can be justified in an
approximate way. By this we mean that even though, strictly speaking,
the $f_{AB}$ are non-singular and one should find a geometrical
interpretation directly for the complete set of them, nevertheless for
some particular regime (energy scale) the $f_{AB}$ can be considered
to be approximately singular and an elimination of spurious degrees of
freedom appropriate. The approximate geometric structure found by this
procedure, though extremely useful, can be completely different from
the exact geometric structure based on the exact characteristics.  We
have illustrated this point with the BEC analog model. The exact
underlying geometric structure provided by the non-relativistic
Gross--Pitaevskii equation is ``parabolic'' while the approximate
effective geometric structure is Lorentzian or ``hyperbolic''.

In this paper we have discussed the different possibilities one can
encounter in extracting a geometric structure from a system of PDEs.
The particular systems that fulfill the Einstein Equivalence Principle
can most easily be seen as analog models {\it of} General Relativity.
Without additional structure they should not be thought of as models
{\it for} general relativity, as the dynamics of the effective
geometry can differ greatly from the proper general relativistic
dynamics (the Einstein equations).  On the other hand, in a previous
work~\cite{normal-modes}, and the in the context of a single field
system, we showed that one-loop quantum effects could also provide, in
some circumstances, a dynamics somewhat resembling that of general
relativity.  We leave as a future project to analyze in detail this
quantum mechanism in the context of a multi-field system.

In summary: The occurrence of effective metrics and effective
geometries in low-energy linearized approximations to wide classes of
dynamical systems is striking --- the near ubiquitous occurrence of
this effect is particularly useful when considering analog models
{\emph{of}} general relativity, can even be interpreted as being
strongly suggestive that some form of ``induced gravity'' may underly
true physical gravity.  To turn ``induced gravity'' into a serious
contender several additional conditions must first be met: On the one
hand we will need to be dealing with multiple background fields (at
least six) in order to have an algebraically general spacetime
metric~\cite{normal-modes}, on the other hand all the linearized
background fields must see the ``same'' Lorentzian metric (either
exactly the same metric if you believe strongly in the Einstein
Equivalence Principle, or at the very least approximately the same
metric in order to be compatible with experimental constraints from
\Eotvos-type experiments). Only once these two basic conditions have
been met will it be sensible to develop a Sakharov-like ``induced
gravity''~\cite{Sakharov}. One-loop effects will then generate a term
in the effective action that is proportional to the Einstein--Hilbert
action~\cite{normal-modes}. The result would then be an ``effective''
theory of gravity in the sense of Donoghue~\cite{Donoghue}, but would
still suffer from the potential defects common to all embedding models
of general relativity~\cite{normal-modes,Regge,Deser}.

In this regard the results we report in this paper are mixed: The good
news is that the occurrence of Lorentzian metrics seems quite common,
the bad news is that {\emph{multiple}} Lorentzian metrics (and worse)
seems generic. We have been able to say quite a bit about the general
way in which one might probe the causal structure of these theories,
and while we have been able to put much of the discussion of
multi-refringence in general framework. We are continuing to work on
seeking field theories that are more general than simple ``minimum
substitution'' theories, but are still restricted enough to produce a
well-controlled metric structure.

\section*{Acknowledgements}

The research of Matt Visser was supported by the US Department of
Energy.  Stefano Liberati was supported by the US National Science
Foundation.  Carlos {\Barcelo} was initially supported by the Spanish
Ministry of Science and Technology (MCYT), and is currently supported
by a European Community Marie Curie grant.  The authors wish to thank
Sebastiano Sonego for his comments and observations.



\begin{thebibliography}{99}
\bibitem{normal-modes}
C. \Barcelo, S. Liberati, and M. Visser,
``Analog gravity from field theory normal modes?'',
Classical and Quantum Gravity {\bf 18} (2001) 3595-3610 [gr-qc/0104001]; 
\\
``Einstein gravity as an emergent phenomenon?'',
gr-qc/0106002; International Journal of Modern Physics (in press).

\bibitem{Courant}
R.~Courant and D.~Hilbert, 
{\em Methods of Mathematical Physics, Vol II}, 
Wiley, John and Sons, (1990).

\bibitem{EDM}
{\em Encyclopedic Dictionary of Mathematics},  
K.~Ito, K.~Ito and N.S.~Ugakkai (Editors), 2nd ed, MIT Press (1987).

\bibitem{Born-Wolf}
M.~Born and E.~Wolf,
{\em Principles of optics},
(Pergamon, Oxford, 1980).

\bibitem{Landau}
L.~D.~Landau, E.~M.~Lifshitz and L.~P.~Pitaevskii, 
{\em Electrodynamics of Continuous Media\/}, 2nd edition (Oxford,
Pergamon Press, 1984).

\bibitem{Hadamard}
J.~Hadamard, 
{\em Le\c{c}ons sur la propagation des ondes et les 
\'equations de l'hydrodynamique}, 
(Hermann, Paris, 1903).

\bibitem{Synge}
J.~L.~Synge,
{\em Relativity: the general theory},
(North Holland, Amsterdam, 1960). 

\bibitem{Asanov}
G.S. Asanov,
{\em Finsler geometry, relativity, and gauge theories},
(Reidel, Dordrecht, Netherlands, 1985).

\bibitem{Bejancu}
A. Bejancu, 
{\em Finsler geometry and applications}, 
(Ellis Horwood, Chichester, England, 1990).

\bibitem{Chern}
D. Bao, S.S. Chern, and Z. Shen,
{\em An introduction to Riemann--Finsler geometry},
(Springer, New York, 2000).

\bibitem{Rund}
H. Rund,
{\em The differential geometry of Finsler spaces},
(Springer, Berlin, 1959).

\bibitem{Antonelli}
P.L. Antonelli and B.C. Lackey (editors),
{\em The theory of Finslerian Laplacians and applications},
(Kluwer, Dordrecht, 1998).

\bibitem{Chern2}
D. Bao, S.S. Chern, and Z. Shen (editors), 
{\em Finsler Geometry},
Proceedings of the Joint Summer Research Conference on Finsler
Geometry, July 1995, Seattle, Washington,
(American Mathematical Society, Providence, Rhode Island, 1996).

\bibitem{Barcelo}
C.~\Barcelo, S.~Liberati, and M.~Visser,
``Analog gravity from Bose--Einstein condensates'',
Class.~Quant.~Grav.~ {\bf 18} (2001) 1137
[gr-qc/0011026];
\\
``Towards the observation of Hawking radiation in 
Bose--Einstein condensates'',
arXiv:gr-qc/0110036.

\bibitem{Garay}
L.~J.~Garay, J.~R.~Anglin, J.~I.~Cirac and P.~Zoller,
``Black holes in Bose--Einstein condensates'',
Phys.~Rev.~Lett.~{\bf 85}, 4643 (2000)
[gr-qc/0002015];\\
``Sonic black holes in dilute Bose--Einstein condensates'',
Phys.~Rev.~ {\bf A63}, 023611 (2001)
[gr-qc/0005131].

\bibitem{cpt01}
M.~Visser, C.~\Barcelo, and S.~Liberati,
``Acoustics in Bose-Einstein condensates as an example of 
broken Lorentz symmetry'',
[hep-th/0109033].

\bibitem{Jackiw}
R.~Jackiw and A.~P.~Polychronakos,
``Supersymmetric fluid mechanics'',
Phys.~Rev.~D {\bf 62}, 085019 (2000).

\bibitem{Raifeartaigh}
L.~O'Raifeartaigh and V.~V.~Sreedhar,
``The maximal kinematical invariance group of fluid dynamics and  
explosion-implosion duality'',
[hep-th/0007199].

\bibitem{Unruh}
W.G. Unruh, 
``Experimental black hole evaporation?'',
Phys. Rev. Lett. {\bf 46}, 1351 (1981);
\\
``Dumb holes and the effects of high frequencies on black hole 
evaporation'',
Phys.~Rev.~D {\bf 51}, 2827 (1995) 
[gr-qc/9409008].

\bibitem{Visser}
M. Visser,
``Acoustic black holes: Horizons, ergospheres, and Hawking radiation'',
Class.~Quantum Grav.~ {\bf 15}, 1767 (1998)
[gr-qc/9712010];\\
``Acoustic propagation in fluids: 
An Unexpected example of Lorentzian geometry'',
gr-qc/9311028;\\
``Acoustic black holes'',
gr-qc/9901047.

\bibitem{Madelung}
E. Madelung, 
``Quantentheorie in hydrodynamischer Form'',
Zeitschrift f\"ur Physik {\bf 38}, 322 (1926).

\bibitem{Bohm}
D. Bohm,
``A suggested interpretation of the quantum theory in terms of 
``hidden'' variables: I and II'',
Phys. Rev. {\bf 85}, 166 (1952);  {\bf 85}, 180 (1952).

\bibitem{Holland}
P.~R.~Holland,
{\em The quantum theory of motion: an account of the de Broglie-Bohm 
causal interpretation of quantum mechanics},
(Cambridge, England, 1993)

\bibitem{Sakharov}
A.~Sakharov, 
``Vacuum quantum fluctuations in curved space and the theory of gravitation'', 
Soviet Physics Doklady, {\bf 12}, 1040 (1968). 
[Dokl.~Akad.~Nauk Ser.~Fiz.~{\bf 177} (1968) 1040].

\bibitem{Donoghue}
J.~F.~Donoghue,
``Quantum general relativity is an effective field theory'',
PRINT-96-198 (MASS.U.,AMHERST)
{\it Talk presented at the 10th International Conference on Problems of 
Quantum Field Theory, Alushta, Ukraine, 13-17 May 1996};
\\
``Gravity and Effective Field Theory: A Talk for Phenomenologists'',
hep-ph/9512287;
\\
``Introduction to the Effective Field Theory Description of Gravity'',
gr-qc/9512024;
\\
``The Ideas of gravitational effective field theory'',
hep-th/9409143;
\\
``General relativity as an effective field theory: 
The leading quantum corrections'',
Phys.~Rev.~D {\bf 50} (1994) 3874
[gr-qc/9405057].

\bibitem{Regge}
T.~ Regge and C.~Teitelboim,
``General Relativity a la string: a progress report'',
in {\em Proceedings of the Marcel Grossman meeting}, Trieste (1975).

\bibitem{Deser}
S.~Deser, F.A.E.~Pirani and D.C.~Robinson,
``New embedding model of general relativity'',
Phys.~Rev.~ {\bf D14}, 3301 1976).

\end{thebibliography}
\end{document}